Title: Letter for the pressure effects on the electrostatic polarizability and dispersive binding
    of off-center polarons in ionic and molecular systems
Author: Mladen Georgiev (Institute of Solid State Physics, Bulgarian Academy of Sciences,
    1784 Sofia, Bulgaria)
Comments: 5 pages pdf format
Subj-class: physics

The question is raised for the pressure effects on the polarizability and dispersive binding in ionic and molecular systems. As an example we take the effect of hydrostatic pressure on the energy gap of phonon coupled two-level systems. Assuming a positive pressure coefficient for the gap, we predict a gradual transition from small vibronic polarons to large polarons as the pressure is increased at constant coupling energy of a two-level system. Concomitantly, the universal dispersive binding degrades from strong at low pressure to weak at high pressure as the system evolves from soft condensed matter to hard condensed matter. Implications are discussed on the development of stellar substance at the early stages of the Universe.

1.  Rationale

Traditionally, a vibronic polaron is the entity which forms as an electron charge carrier couples to a pseudo-Jahn-Teller distortion in ionic or molecular crystals [1]. In the simplest case of a two-level system, the distortion itself arises as the phonon mode couples to mix the two electronic states involved. In so far as the two levels nearly-degenerate initially, an electronic hybrid forms which breaks down the original symmetry at the nonmixing configuration. For a two-level system with initial inversion symmetry, this implies parity nonconservation and the related occurrence of an electric dipole [2]. The correlated configurational distortion pattern is asymmetric to evidence the fact that parity-breaking has been involved.

Regarding the energy gap in a two-level system, the vibronic polaron is small if the gap is largely inferior to four times the coupling energy (Janh-Teller energy, to be exact) and large if the gap is close though still inferior to that amount. The coupling energy being characteristic of the inverse distortion radius [3], the terms "small" and "large" simply reflect the relative polaron size in a crystalline medium. Certainly, the external hallmark of the polaron size is the magnitude of the inversion electric dipole which is ultimate if the vibronic polaron is small and is tending to vanish as its size increases.

The off-center polarons are among the most thrilling objects of interest for solid state physics. Besides their tunneling transitions across the central barrier, they perform barrier-hindered rotations about the central site [4]. The quantum mechanical entity is conceived smeared about the central barrier and along the orientational barrier path. As a result, its shape may remind the oblate form (symmetry axis and rotational axis of unit are perpendicular to each other) and prolate form (symmetry axis and rotational axis of unit are parallel) depending on the coupled phonon modes symmetries. These analogies will be discussed in greater detail elsewhere.

Now imagine the complex entity of vibronic polaron to be subjected to a hydrostatic pressure. There are several options for a pressure effect on the polaron parameters, such as the initial energy gap, the coupling energy, the coupled vibrational frequency, and/or the polaron size. To begin with, we first choose the gap. Energy bandwidths and gaps in most semiconductors are known to widen as the hydrostatic pressure is increased [5]. Following that logic, we see that the polaron will evolve from small to large as the pressure is increased. Its hallmark will in turn decrease and so will its polarizability and dispersive binding with neighboring units. What is important is that the vibronic system will evolve from strongly bound to weakly bound, as the pressure is increased, e.g. from the one in soft condensed matter to one in hard condensed matter. In many ways soft condensed matter behaves like a molecular crystal [6]. We now have a *motif* to pursue the pressure problem in greater detail.

## 2. Off-center polaron Hamiltonian

Herein we address the off-center polaron problem by second quantization in which the phonon coordinate is regarded as a c-number [7]. The Hamiltonian reads

$$H = \sum_{m\alpha} t_{m\alpha} a_{m\alpha}^\dagger (a_{m+1\alpha} + a_{m-1\alpha}) + \sum_{m\alpha} E_{m\alpha} a_{m\alpha}^\dagger a_{m\alpha} +$$

$$\sum_{i\alpha\beta} G_{i\alpha\beta} q_{i\alpha\beta} (a_{m\alpha}^\dagger a_{m\beta} + a_{m\beta}^\dagger a_{m\alpha}) + \tfrac{1}{2} \sum_{i\alpha\beta} K_{i\alpha\beta} q_{i\alpha\beta}^2 , \qquad (1)$$

where $K_{i\alpha\beta} = M_{i\alpha\beta} \omega_{i\alpha\beta}^2$ is the stiffness. Minimizing H in $q_{i\alpha\beta}$ results in

$$q_{i\alpha\beta 0} = -(G_{i\alpha\beta}/K_{i\alpha\beta})(a_{m\alpha}^\dagger a_{m\beta} + a_{m\beta}^\dagger a_{m\alpha}) \qquad (2)$$

which is inserted back to (1) to give (adiabatic exclusion):

$$H_{min} = \sum_{m\alpha} t_{m\alpha} a_{m\alpha}^\dagger (a_{m+1\alpha} + a_{m-1\alpha}) +$$

$$\sum_{m\alpha} E_{m\alpha} a_{m\alpha}^\dagger a_{m\alpha} - \sum_{i\alpha\beta} E_{JTi\alpha\beta} (a_{m\alpha}^\dagger a_{m\beta} + a_{m\beta}^\dagger a_{m\alpha})^2 \qquad (3)$$

where we have introduced the Jahn-Teller energies $E_{JTi\alpha\beta} = G_{i\alpha\beta}^2 / 2K_{i\alpha\beta}$ [1,2]. Our further interest will be focused onto the local Hamiltonian $H_{minloc}$ composed of terms #2 and #3 in (3) default of its #1 hopping term. We solve for the Schrodinger equation $H_{min}\Psi = E\Psi$ by a linear combination of one-particle one-band states $\Psi = c_\alpha a_\alpha^\dagger |0> + c_\beta a_\beta^\dagger |0>$ (site index omitted) where $|0>$ is the vacuum state, $\alpha$ and $\beta$ are two narrow electron bands ($\approx$ two levels). Inserting we get the averages:

$$<a_\alpha|H_{minloc}|a_\alpha^\dagger> = E_\alpha - E_{JT\alpha\beta} <a_\alpha|(a_\alpha^\dagger a_\beta + a_\beta^\dagger a_\alpha)^2|a_\alpha^\dagger> = E_\alpha$$

$$<a_\beta|H_{minloc}|a_\beta^\dagger> = E_\beta - E_{JT\alpha\beta} <a_\beta|(a_\alpha^\dagger a_\beta + a_\beta^\dagger a_\alpha)^2|a_\beta^\dagger> = E_\beta$$

$$<a_\beta|H_{minloc}|a_\alpha^\dagger> = -E_{JT\alpha\beta} <a_\beta|(a_\alpha^\dagger a_\beta + a_\beta^\dagger a_\alpha)^2|a_\alpha^\dagger> = -E_{JT\alpha\beta} \qquad (4)$$

while the secular equation for the energy is

$$(<a_\alpha|H_{minloc}|a_\alpha^\dagger> - E)(<a_\beta|H_{minloc}|a_\beta^\dagger> - E) -$$

$$<a_\beta|H_{minloc}|a_\alpha^\dagger><a_\alpha|H_{minloc}|a_\beta^\dagger> = 0$$

with two roots reading

$$E_\pm = \tfrac{1}{2}\{(E_\alpha + E_\beta) \pm \sqrt{[(E_{gap\alpha\beta})^2 + E_{JT\alpha\beta}^2]}\} \qquad (5)$$

where $E_{gap\alpha\beta} = |E_\alpha - E_\beta|$ is the interband (interlevel) energy gap.

Equation (5) gives extremal electron energies pertaining to the bottom of the potential well. Otherwise, the dependence of the electron energy on the coupled phonon coordinate $q_{i\alpha\beta}$ is obtained from the nonextremal Hamiltonian as in (3)

$$H_{loc} = \sum_{m\alpha} E_{m\alpha}\, a_{m\alpha}^\dagger a_{m\alpha} + \sum_{i\alpha\beta} G_{i\alpha\beta} q_{i\alpha\beta}(a_{m\alpha}^\dagger a_{m\beta} + a_{m\beta}^\dagger a_{m\alpha}) + \tfrac{1}{2}\sum_{i\alpha\beta} K_{i\alpha\beta} q_{i\alpha\beta}^2$$

where the coupled-phonon (vibronic) Hamiltonian is

$$H_{phon} = \sum_{m\alpha} E_{m\alpha}\, a_{m\alpha}^\dagger a_{m\alpha} + G_{i\alpha\beta} q_{i\alpha\beta}(a_{m\alpha}^\dagger a_{m\beta} + a_{m\beta}^\dagger a_{m\alpha}) + \tfrac{1}{2} K_{i\alpha\beta} q_{i\alpha\beta}^2 \qquad (6)$$

This is the Hamiltonian of a displaced harmonic oscillator moving in a double well potential [7]. The bottom positions of the wells are located at $\pm q_{\alpha\beta 0}$, as given by eqn. (2). The exact solution for the double-well oscillator not being available, we use a linear combination of the eigenstates for the left-hand and right-hand wells. These eigenstates overlap as $\langle \chi(x+q_0) | \chi(x-q_0)\rangle = \exp(-\xi_{0\alpha\beta}^2)$ to give a tunneling splitting of magnitude

$$\Delta\tau_{sp} = E_{gap\alpha\beta} \exp(-\xi_{\alpha\beta 0}^2),$$

$$\xi_{\alpha\beta 0} = \sqrt{(M_{\alpha\beta}\omega_{\alpha\beta}^2/\eta\omega_{\alpha\beta})}\, q_{\alpha\beta 0}, \quad q_{i\alpha\beta 0} = -(G_{i\alpha\beta}/K_{i\alpha\beta})(a_{m\alpha}^\dagger a_{m\beta} + a_{m\beta}^\dagger a_{m\alpha}) \qquad (7)$$

is the mode coordinate in dimensionless units ($\eta = h/2\pi$). We note that all the above reasoning is holding good for $4E_{JT} \gg E_{gap}$ which is the small-polaron condition. We now see $\Delta\tau_{sp} \ll E_{gap}$.

The large-polaron condition is $4E_{JT} \geq E_{gap}$. The tunneling splitting is still given as above but in this case care should be taken in so far as the vibronic level is closer to the barrier top. Now the exponent in (7) is small and expanding in a power series gives $\exp(-\xi_{\alpha\beta 0}^2) \approx 1 - \xi_{\alpha\beta 0}^2$ to result in

$$\Delta\tau_{lp} \approx E_{gap\alpha\beta}(1 - \xi_{\alpha\beta 0}^2) = E_{gap\alpha\beta}[1 - \sqrt{(M_{\alpha\beta}\omega_{\alpha\beta}/\eta)}(G_{\alpha\beta}/K_{\alpha\beta})(a_{m\alpha}^\dagger a_{m\beta} + a_{m\beta}^\dagger a_{m\alpha})] \qquad (8)$$

We now get $\Delta\tau_{lp} \leq E_{gap}$.

### 3. Hydrostatic pressure

There have been a number of reported cases in the periodic literature as well as in some monographs of the pressure effects on the electronic bandwidths and bandgaps in solids [5]. Both hydrostatic and uniaxial pressures have been applied. We start with the former. In most cases the pressure-dependence coefficient $dE/d\pi > 0$, so that the band quantity increases with the hydrostatic pressure $\pi$.

In the present narrow band approach the only significant band quantity is the interlevel (band) gap $E_{gap}$. To estimate the effect of hydrostatic pressure on the bandgap $E_{gap}$ and the tunneling splitting

$$\Delta\tau = E_{gap}\exp(-\xi_0^2), \qquad (9)$$

we define $dE_{gap}/d\pi = \kappa > 0$ to obtain

$$\Delta\tau(\pi) - \Delta\tau(\pi_0) = \kappa\exp(-\xi_0^2)(\pi - \pi_0) = [E_{gap}(\pi) - E_{gap}(\pi_0)]\exp(-\xi_0^2) > 0 \text{ for } \pi > \pi_0 \qquad (10)$$

We see that the ground-state tunneling splitting increases as the pressure is raised, the polaron eventually evolving from small to large.

### 4. Dispersive coupling

Following F. London's now classic description [8], we consider the dispersive coupling which arises between polarizable atoms and/or vibronic polarons, due to the interaction between fluctuating dipoles. The coupling energy is

$$U_{VdW} = -\tfrac{1}{2} (\Delta\tau) [\alpha(T)/\varepsilon R^3]^2 \qquad (11)$$

where

$$\alpha(T) = [p_{12}^2/(\Delta\tau)] f(T) \qquad (12)$$

is the vibronic polarizability in which $f(T)$ is a definite function of temperature depending on the dimensionality of the problem, $p_{12}$ is an electric dipole, $\varepsilon$ is a dielectric constant [11]. Inserting into (11) we obtain

$$U_{VdW} = -\tfrac{1}{2} [p_{12}^4 f(T)^2/\varepsilon^2 R^6] /(\Delta\tau) \qquad (13)$$

We see that the vibronic tunneling splitting $\Delta\tau$ being small (usually of the order of a few μeV to 1 meV), gives an enormous strength to the polarizability of a two-level system resulting in a colossal dispersive (or Van der Waals) force. Namely, inserting (9) into (13) we get

$$U_{VdW} = -\tfrac{1}{2} \{[p_{12}^4 f(T)^2/\varepsilon^2 R^6] / E_{gap}\}\exp(+\xi_0^2) \qquad (14)$$

which is exponentially large. $E_{gap}$ increasing with the pressure $\pi$, the dispersive energy $U_{VdW}$ drops down as the hydrostatic pressure is increased. The reason for this behavior is that the polaron turns from small to large which is only weakly coupled.

### 5. Molecular binding in gaseous and condensed systems

We imagine a gaseous or soft condensed matter system under low pressure due to low atom concentration. The energy gap $E_{gap}$ of the two-level system being small, so will the ground-state tunneling splitting (9) and the system belongs vibronically to the small-polaron range. As we have seen, this range is characterized by a colossal VdW attraction by eqns.(13) or (14) extending over many atomic units. Due to this attraction, there will always be a tendency for cluster formation in the VdW field which if materialized will produce loose aggregates of constituent atoms. As the pressure *within* an aggregate increases on its being a solid pre-particle, the energy gap $E_{gap}$ will increase and, concomitantly, the tunneling splitting by eqn.(9) will also increase gradually leaving the small-polaron range to become large polaron. According to eqns. (13) and (14) the VdW binding will be gradually pressure-weakened to the extent that wider-spaced pairs may not withstand thermal agitation to decompose. Some pairs will undoubtedly survive to give rise to a new solid phase. The final result is likely to be a planet formation within a gas of constituent atoms. There are so many of these atoms within an enormous volume of free gas! The remaining VdW coupling within a planet will be just moderate in strength, comparable to what we usually encounter here on Earth. At this stage, other interactions may prevail inside the planet giving stronger cohesion but the original role of the VdW coupling will be to trigger the nucleation process whatever its nature because of the universality of the VdW bond.

There has long been a search among cosmology chemists for an universal attraction which may bring the right atoms together to form the appropriate chemical compounds. Some see the hand of an Intelligent Design in selecting the appropriate surrogates [9], and if so, the long range (colossal) VdW force might be His tool. At this stage the colossal force is only conceivable but there are definite indications that it plays a role in forming samples of unusual matter, e.g. in atmospheric discharges where large amounts of photoexcited atoms are born and gain VdW coupling, perhaps as at the early stages of the Universe.

## 6. Conclusion

A logical conclusion of the present paper is the breaking down of inversion symmetry and the related appearance of an inversion electric dipole which gives rise to a number of important implications, such as the occurrence of electrostatic VdW interactions of immense vibronic magnitude in dilute systems. We stress on electrostatics since a similar theory can be devised of magnetic coupling leading to a magnetostatic analogue of the electrostatic binding. It may be the subject of our further attention.